# Abrupt switching of the anomalous Hall effect by field-rotation in nonmagnetic ZrTe$_5$

Joshua Mutch[1], Xuetao Ma[2], Chong Wang[3], Paul Malinowski[1], Joss Ayres-Sims[1], Qianni Jiang[1], Zhoayu Liu[1], Di Xiao[3], Matthew Yankowitz[1,2#], Jiun-Haw Chu[1#]

[1]Department of Physics, University of Washington, Seattle, WA 98105, USA
[2]Department of Material Science and Engineering, University of Washington, Seattle, WA 98105, USA
[3]Department of Physics, Carnegie Mellon University, Pittsburg, PA 15213, USA

# correspondence to jhchu@uw.edu; myank@uw.edu

**Abstract: The Hall effect arises when time reversal symmetry is broken by either intrinsic magnetism or an external magnetic field[1]. The latter contribution dominates in non-magnetic materials, in which the angular dependence of the Hall effect is typically a smooth cosine function because only the out-of-plane projection of the field generates the in-plane transverse motion of electrons. Here, we report the observation of an abrupt switching of the Hall effect by field rotation in a non-magnetic material, ZrTe$_5$. The angular dependence of the Hall resistivity approaches a signum function, persisting down to an extremely low field of 0.03 T. By varying the carrier density of ZrTe$_5$ over three orders of magnitude, we show that this singular behavior is due to the anomalous Hall effect generated by the ultra-dilute massive Dirac carriers in the quantum limit of Pauli paramagnetism when the Zeeman energy exceeds the Fermi energy. Our results elucidate the origin of the anomalous Hall effect in ZrTe$_5$, arising owing to the spin-polarized massive Dirac electrons rather than the separation of Weyl nodes.**

**Main text:**

The transition metal pentatelluride ZrTe$_5$ presents a unique paradigm for three-dimensional (3D) topological materials. Early measurements revealed a Dirac-like bulk electronic structure with a Fermi velocity comparable to graphene[2-4]. Unlike graphene, however, ZrTe$_5$ is not a topological semimetal because the band touching at the Dirac point is not explicitly protected by any crystal symmetry. Instead, the Dirac-like dispersion of ZrTe$_5$ emerges from its proximity to the phase boundary between strong and weak topological insulator phases. The magnitude of the Dirac point gap grows as the system is pushed further away from the topological phase boundary[5,6], endowing the band edge carriers with a mass term[7-9]. Since the massive Dirac band is the only band crossing the Fermi level at low carrier density, ZrTe$_5$ serves as a model platform to explore the physics of relativistic quasiparticles in a 3D system.

The unusual electronic structure of ZrTe$_5$ has already enabled a series of striking discoveries, including the chiral magnetic effect[4], log-periodic oscillations[10], a 3D quantum Hall effect[11], and the anomalous Hall effect (AHE)[12,13]. The observation of the AHE is particularly intriguing, given that ZrTe$_5$ is not an intrinsically magnetic material. The AHE has previously been argued to arise from Weyl crossings induced by the external magnetic field[12], in which a pair of Weyl nodes act as the source and sink of Berry flux[14]. More recent measurements suggest that an alternative mechanism may instead be responsible, in which the large Berry curvature driving the AHE arises from an avoided crossing of the massive Dirac bands, without any Weyl node formation[13,15]. Despite considerable prior experimental attention, the origin of the AHE in ZrTe$_5$ remains to be fully understood.

In this work, we present extensive measurements of the AHE in both bulk single crystals and exfoliated thin flakes of ZrTe$_5$. We find that the Hall conductivity of all samples can be precisely separated into two components, corresponding to the anomalous and orbital Hall contributions. The carrier density varies by over three orders of magnitude in our samples, enabling a quantitative analysis of the AHE as a function of doping. Our measurements unambiguously identify the spin-splitting of the massive Dirac bands as the origin of the AHE in ZrTe$_5$, and are incompatible with the formation of Weyl nodes. Furthermore, we show that ZrTe$_5$ enters a spin-polarized quantum limit state at extremely low magnetic field, giving rise to a singular angle dependence of Hall response reminiscent of a uniaxial ferromagnet.

ZrTe$_5$ has a van der Waals (vdW) layered structure with the Cmcm orthorhombic space group. Each layer consists of ZrTechains extending along the *a*-lattice direction, and the layers are stacked along the *b*-lattice direction (Fig. 1a). For convenience we will adopt the coordination system, in which the *x, y* and *z*-axes correspond to the lattice *a, c* and *b*-axes respectively. Figures 1b-d show the field dependence of resistivity and conductivity tensors of bulk ZrTe$_5$. We see a cusp-like behavior in measurements of $\rho_{xx}$ versus field (Fig. 1b), saturating to a large magnetoresistance (~400%) at a relatively low field (0.2 T). The saturating behavior is consistent with the magnetoresistance of a single band system with a closed semi-classical orbit[16]. $\rho_{xy}$ exhibits a nonlinear field dependence, which has previously been interpreted in the context of either a two-band model[10] or the AHE[12,13]. We obtain the field dependence of the longitudinal $\sigma_{xx}$ (Fig. 1c) and the Hall conductivity $\sigma_{xy}$ (Fig. 1c-d) by inverting the resistivity tensor (see methods). The Hall conductivity shows a dispersive resonant-like behavior, in which $\sigma_{xy}$ rapidly reaches a peak value at a very low field (0.01 T) and decreases gradually afterwards. Such a field dependence of $\sigma_{xy}$ corresponds to the single-band Drude orbital Hall conductivity

$$\sigma_{xy}^{orbital} = \frac{pe\mu^2 B}{1+\mu^2 B^2}$$

where $p$ is the carrier density and $e$ is the electron charge. In this expression, $\sigma_{xy}^{orbital}$ displays a maximum at a peak field corresponding to $\mu B = 1$, providing a direct measure of the average mobility, $\mu$. In systems with highly mobile carriers, such as the Dirac semimetal Cd$_3$As$_2$, the condition $\mu B = 1$ can be achieved at a very low field[17]. The sharp peak of $\sigma_{xy}$ at 0.01T in Fig. 1d indicates that the mobility of our sample is also extremely high. However, whereas the low-field behavior of $\sigma_{xy}$ is well described by $\sigma_{xy}^{orbital}$, the high-field behavior shows a clear deviation. Instead of decaying to zero as expected from the expression for $\sigma_{xy}^{orbital}$, our measurements of $\sigma_{xy}$ in ZrTe$_5$ saturate to a finite value in high field (see Extended Data Fig. 6 for more samples and higher fields). This contrasts previous observations in Cd$_3$As$_2$, in which $\sigma_{xy}$ exhibits the anticipated decay to zero at high field[17].

We decompose our measured $\sigma_{xy}$ in Fig. 1d by fitting $\sigma_{xy} = \sigma_{xy}^{orbital} + \sigma_{xy}^{AHE}$ where $\sigma_{xy}^{AHE} = \sigma_0^{AHE} \tanh B/B_0$. The choice of a hyperbolic tangent function for the $\sigma_{xy}^{AHE}$ component is purely phenomenological in order to capture the step function behavior, and the sharpness of the step function is characterized by the saturation field $B_0$. We find $\sigma_{xy}^{AHE}$ saturates to 0.06 S/cm with a saturation field of $B_0 \sim 0.02$ T. The data cannot be fit by a two-band orbital model, ruling out this scenario (Extended Data Fig. 7). From the $\sigma_{xy}^{orbital}$ fit, we extract the carrier density, $p \sim 2.2 \times 10^{12}$ cm$^{-3}$, and average mobility, $\mu = 1.7 \times 10^6$ cm$^2$/(V·s). This high mobility is comparable to the cleanest Dirac and Weyl semimetals previously investigated, including Cd$_3$As$_2$[17] and TaAs[18].

Having established the decomposition of $\sigma_{xy}$ to the $\sigma_{xy}^{orbital}$ and $\sigma_{xy}^{AHE}$ components, we now examine how each component varies as a function of carrier density. We utilize three methods to tune the carrier density of ZrTe$_5$ by more than three orders of magnitude (from $p \sim 9.8 \times 10^{12}$ cm$^{-3}$ to $p \sim 7.4 \times 10^{15}$ cm$^{-3}$). First, we purify the starting materials in the crystal growth process (see Methods for more details), which substantially reduces the carrier density. Second, we mechanically exfoliate bulk crystals down to flakes with thicknesses ranging from $105 - 350$ nm. The reduction of the sample thickness appears to increase the carrier density[19]. Third, we utilize a parallel plate capacitor geometry to apply a back-gate voltage to the 105 nm thick exfoliated sample resting on a Si/SiO$_2$ wafer. Although gating a thick sample inevitably introduces a vertical potential gradient, we nevertheless observe a monotonic modulation of the carrier density, as determined from the fitting of $\sigma_{xy}^{orbital}$, and from a gate-dependent change in the resistivity. We are able to achieve a 50% change in the carrier density by applying 80 V to the Si back gate with a 285 nm capping layer of SiO$_2$ (Extended Data Figure 9). Although the origin of this carrier density variation in thin flakes and the mechanism of gating are not fully understood, it offers a convenient tuning knob to tune the relative contribution of to $\sigma_{xy}^{orbital}$ and $\sigma_{xy}^{AHE}$. As we discuss below, these tuning capabilities are the key to understanding the origin of AHE.

For all the samples, $\sigma_{xy}$ can be cleanly decomposed into $\sigma_{xy}^{orbital}$ and $\sigma_{xy}^{AHE}$ according to the methodology shown in Fig. 1d. Figure 2a shows the carrier density ($p$) and the mobility ($\mu$) extracted from the $\sigma_{xy}^{orbital}$ component for all of our measured bulk single-crystals and exfoliated flakes. In Fig. 2b-c we present the saturation value $\sigma_0^{AHE}$ and the saturation field $B_0$. All three quantities $\mu$, $\sigma_0^{AHE}$ and $B_0$ exhibit clear trends as a function of $p$. The mobility $\mu$ increases with reducing carrier density without any sign of a metal-insulator transition, characteristic of narrow or zero-gap semiconductors[20]. On the other hand, both $\sigma_0^{AHE}$ and $B_0$ grow with carrier density, increasing from $\sigma_0^{AHE} = 0.95$ S/cm and $B_0 = 0.01$ T for the bulk crystal with the most dilute carriers ($p = 9.8 \times 10^{12}$) to $\sigma_0^{AHE} = 34$ S/cm and $B_0 = 0.16$ T for the 105nm thick exfoliated sample ($p = 7.4 \times 10^{15}$ cm$^{-3}$).

Currently, there are two competing explanations for the origin of the AHE in ZrTe$_5$. In one scenario, Weyl nodes form owing to an inversion of the valence and conduction bands in an external magnetic field[12]. The AHE results from the finite Berry curvature of the Weyl nodes when the Zeeman splitting, $E_Z$, exceeds the energy gap, $\Delta$. The AH conductivity grows continuously subsequent to Weyl node formation since it is proportional to their separation in momentum space, before eventually saturating at large $B$ when the two Weyl points annihilate each other at the boundary of Brillouin zone. In the other scenario, $E_Z < \Delta$ and hence the valence and conduction bands stay well separated and Weyl nodes do not form. Although the massive Dirac bands are endowed with very large Berry curvature at the band edge, the net Berry curvature remains zero at $B = 0$ since time reversal symmetry (TRS) enforces a degeneracy between bands with opposite spin (and therefore opposite signs of the Berry curvature). However, the Zeeman splitting of the bands exposes a finite Berry curvature at the band extrema when TRS is broken by a magnetic field. The AHE onsets immediately upon lifting the spin degeneracy, and then saturates at the quantum limit when the $E_Z$ exceeds the Fermi energy, $E_F$, after which all the electrons are spin-polarized. Fig. 2d shows an illustration of this scenario.

These competing scenarios present distinct, experimentally testable predictions for the behavior of the AHE for magnetic fields aligned in various directions with respect to the principal crystal axes of ZrTe$_5$. For the purposes of this discussion, we take band structure parameters and g-factors previously reported for ZrTe$_5$, however the conclusions of our analysis do not depend on the exact values of these parameters (see Methods). Assuming $\Delta = 10$ meV, $g_z = 21.3$, and a valence band bandwidth of 150 meV along the interlayer direction[9], we estimate that Weyl nodes form for $B >$ 8 T, with a saturation field of $B_0 \sim 120$ T corresponding to their annihilation. This sharply contrasts with our experimental observations, in which we find that the anomalous Hall conductivity (AHC) is well described by a step function that onsets at extremely small $B > 0$, with a saturation field as low as 0.01 T in bulk crystals, and 0.17 T in the thinnest exfoliated crystals we measured.

In contrast, our observations match well with the predictions of the latter scenario, in which the AHE arises due to spin-split massive Dirac bands with large Berry curvature. We find that $B_0$ increases monotonically with $p$ (Fig. 2c), consistent with the expectation that larger $B$ is required to reach the spin-polarized quantum limit ($E_Z > E_F$) since the chemical potential increases with doping (see schematics in Fig. 2d). Furthermore, the monotonic increase of $\sigma_{xy}^{AHE}$ with carrier density is consistent with the expectation that the Fermi surface encloses more Berry flux generated by the Dirac mass for larger $E_F$.

We calculate the intrinsic AHC using the previously reported $\mathbf{k} \cdot \mathbf{p}$ Hamiltonian of ZrTe$_5$ (Methods). The results reproduce all the qualitative features described above, including the saturation of the AHC when $E_Z > E_F$ (Extended Data Figure 8), and the increase of the AHC with carrier density. We plot the calculated saturation field and the saturated AHC as a function of carrier density $p$ (red dotted lines in Figs. 2b-c). Although the calculated AHC depends on the specific band parameters for a given doping, the doping dependence of the AHC is robust against changes in these parameters and exhibits excellent agreement with our data. Our results reveal a striking picture of ZrTe$_5$: even without intrinsic magnetism, ZrTe$_5$ becomes a spin-polarized metal upon reaching the quantum limit of Pauli paramagnetism in a field as low as 0.01 T. Because of the ultra-high mobility, the orbital Hall effect is strongly suppressed at this weak field (peak field $\sigma_{xy}^{orbital} \sim 0.01$ T), therefore the transverse conduction is dominated by the dissipationless AH current.

Finally, we turn to the angular dependence of the Hall effect, in which we rotate a constant magnetic field either within the *bc* or *ba* plane. As shown in Figs. 3a-b, the magnitude of the Hall resistivity is nearly constant as a function of angle, but switches sign abruptly when the field rotates across the *ac* plane. This signum function angular dependence persists down to an extremely low magnetic field, below which the angular dependence returns to a smooth cosine function. We additionally find that this abrupt switching effect persists up to ~135 K for $B = 14$ T, as shown in Fig. 3c for a representative bulk crystal. This corresponds approximately to the condition where $k_B T > E_Z$, providing additional evidence that spin-splitting of the massive Dirac valence band is required in order to see the abrupt switching of $\rho_{xy}$.

Although previous work has also shown step-function dependence in the AH resistivity[12], this feature is only observed at fields above 1 T and additionally requires a background subtraction to eliminate the orbital Hall component. Here, we find that the raw $\rho_{xy}$ exhibits this singular behavior at B > 0.03 T, enabled by our synthesis of ZrTe$_5$ samples with the most dilute carriers and highest mobility. Singular angle dependence of $\rho_{xy}$ has previously been observed only in uniaxial ferromagnets, in which it arises from the abrupt reversal of the intrinsic magnetization. Owing to the combination of the large anisotropy of the g-factor of ZrTe$_5$ (21.3 along the *b*-axis versus ~2-3.2 in-plane[3]) and the ultra-low carrier density of our samples, a very small amount of out-of-plane field projection is sufficient to drive the system into the spin-polarized quantum limit. The resulting plateau of the AH conductivity as a function of angle can only be understood in this context, as magnetic field-induced Weyl node formation is not expected to result in such abrupt switching at very low fields[15].

The AHE in ZrTe$_5$ is exceptional even in the broader context of magnetic materials. We calculate the anomalous Hall angle, $\theta_{AHE} = \tan^{-1}(\sigma_0^{AHE}/\sigma_{xx})$ for all the samples we measured and plot them against the AHC (Fig. 3d). We find that $\theta_{AHE}$ reaches a value of 17.2° as the carrier density increases, comparable to the magnetic topological semimetals with the largest $\theta_{AHE}$[21-23], albeit with smaller AHC due to low carrier density. This large Hall angle is another signature of dissipationless nature of the anomalous Hall current. Intriguingly, for a high carrier density bulk sample the Hall angle, $\theta_H = \tan^{-1}(\rho_{xy}/\rho_{xx})$, reaches a value as large as 64° in fields as low as 0.04 T without any background subtraction of $\rho_{xy}$ (Extended Data Figure 10). This large Hall angle is likely a result of the combination of both anomalous Hall and orbital Hall effects, and it is approaching the 90° limit of the 3D quantum Hall effect that was also discovered in ZrTe$_5$[11]. Understanding how the AHE in the quantum limit of Pauli paramagnetism evolves towards the 3D QHE remains an open question for future study.

Note added: During preparation of this manuscript, we became aware of a similar study by Liu *et al.*[24]. In this study Liu *et al.* analyzed the Hall conductivity of exfoliated thin flakes of ZrTe$_5$ and HfTe$_5$ with a single doping level. Their analysis and the interpretation broadly corroborate our findings.

**Method:**

Material growth and sample preparation

Single-crystal ZrTe$_5$ was grown with the flux method. The Zr slugs (99.9% pure, Alfa Aesar) and Te shot (99.9999% pure, Alfa Aesar) were loaded into a quartz ampule in Zr:Te ratios of either 1:100 or 1:200. The ampule was warmed to 900°C in 9 hours, kept at 900°C for 72 hours, and then cooled to 505°C in 48 hours. To promote large crystal growth, the ampule was repetitively cooled to 440°C and warmed to 505°C. Last, the ampule was cooled to 460°C and decanted in a centrifuge at this temperature.

To achieve extremely low carrier density in these flux-growths, we found the most important growth parameter was the purity of the starting zirconium and tellurium materials. For all of our growths, 99.9999% pure tellurium was used (Alfa Aesar). Using this tellurium with a standard zirconium source (99.5 or 99.95 slug, Alfa Aesar) leads to carrier densities typically as low as $1 \times 10^{16}$ cm$^{-3}$. To achieve even lower carrier densities, we found it necessary to purify the zirconium source through arc melting. We arc melted our zirconium source three times iteratively. We discovered that growth using arc-melted zirconium yielded crystals with carrier densities as low as $\sim 1 \times 10^{12}$ cm$^{-3}$. The importance of the purity of zirconium is consistent with the much lower purity rating of commercially available zirconium compared to tellurium.

For bulk transport measurements, single crystals of typical dimensions 1.5 mm × 0.1 mm × 0.02 mm were sputtered with gold, and then 25-μm-diameter gold wires were placed on the crystals and adhered with silver paint. The resistance was measured with an SRS830 lock-in amplifier and an SRS CS580 current source. Given the needle-like nature of the single crystals, the resistance was measured along the *a*-axis of the crystal.

For the exfoliated measurements, the bulk ZrTe$_5$ crystals were exfoliated onto Si/SiO$_2$ wafers using the Scotch tape method. Devices were fabricated using standard e-beam lithography and metal deposition of chromium/gold (4/300 nm for flakes 245 nm or less, 5/430 nm for the 350 nm flake).

Images of exfoliated devices and a bulk crystal can be found in Extended Data Fig. 1. A summary of the temperature dependence of resistivity can be found in Extended Data Fig. 2.

Calculation of $\sigma_{xy}$ from resistivity tensor

To properly quantify $\sigma_{xy}$ it is necessary to know $\rho_{xx}$, $\rho_{yy}$, and $\rho_{xy}$. The measurement of $\rho_{yy}$ is difficult due to the needle-like nature of the crystals. We have performed the measurement $\rho_{yy}$ in selected samples (Extended Data Fig. 3 (a)). For the typical field range ($|B| < 0.2T$) used in our measurements, we found that approximating $\rho_{yy} = k\rho_{xx}$ is valid, where $k$ is a constant (Extended Data Fig. 3 (b)). We measured $k \sim 3.5$, slightly higher than previous reports in zero-field[25,26]. Although increasing the value of $k$ does have the impact of reducing both the carrier density $p$ and the anomalous conductivity $\sigma_{xy}^{AHE}$, the overall relationship between $\sigma_{xy}^{AHE}$ and $p$ is not affected (Extended Data Fig. 3 (e)). Consequentially, the main conclusions of this paper are independent of the choice of $k$ value.

Extended Data Fig. 4 shows the measured $\sigma_{xy}$ signal and the fitting components $\sigma_{xy}^{orbital}$ and $\sigma_{xy}^{AHE}$ for all crystals measured. In all cases, the fitting terms $\sigma_{xy}^{orbital} + \sigma_{xy}^{AHE}$ trace the measured signal very well. The extracted fitting parameters are summarized in Extended Data Table 1.

Calculation of anomalous Hall conductivity from the $\boldsymbol{k} \cdot \boldsymbol{p}$ Hamiltonian

We adopt the $\boldsymbol{k} \cdot \boldsymbol{p}$ Hamiltonian up to linear order in $k$ in Ref [27]:

$$H = M\tau_z + v_x k_x \tau_x \sigma_y + v_y k_y \tau_y + v_z k_z \tau_x \sigma_x + Z\sigma_z,$$

where $\tau$ and $\sigma$ are Pauli matrices denoting orbital and spin degrees of freedom, and $Z\sigma_z$ corresponds to the Zeeman energy. The parameters are also from Ref [27]: $M = 7.5$ meV, $v_x = 4.51$ eV·Å, $v_y = 2.70$ eV·Å, $v_z = 0.33$ eV·Å. For each $Z$, we set the Fermi energy at the top of the lower valence band and calculate both anomalous Hall conductivity and carrier density, which are presented in Fig. 2b-c. We also calculate the anomalous Hall conductivity as a function of $Z$ at a fixed carrier density (Extended Data Figure 8).

**Acknowledgements:** We thank Liang Fu, Dima Pesin, Binghai Yan for helpful discussions. This work is primarily supported as part of Programmable Quantum Materials, an Energy Frontier Research Center funded by the U.S. Department of Energy (DOE), Office of Science, Basic Energy Sciences (BES), under award DE-SC0019443. Materials synthesis at UW was partially supported by the Gordon and Betty Moore Foundation's EPiQS Initiative, Grant GBMF6759 to JHC. A portion of this work was performed at the National High Magnetic Field Laboratory, which is supported by the National Science Foundation Cooperative Agreement No. DMR-1644779, the State of Florida. M.Y. and J.H.C. acknowledge the support from the State of Washington funded Clean Energy Institute. J.H.C. also acknowledge the support of the David and Lucile Packard Foundation.

**Author contributions:** J.-H.C., M.Y. and D.X. supervised the project. J.M. and J. A. synthesized bulk crystals. J.M., P.M., Q.J., Z.L. and J. A. measured bulk crystals. J.M. and X.M. fabricated devices and performed the measurements on exfoliated flakes. C.W. and D.X. provided theoretical support. J.M., M.Y. and J.-H.C. wrote the manuscript with input from other authors. All authors discussed results.

**Competing Financial Interests:** The authors declare no competing financial interests.

**Data Availability:** The data that support the findings of this study are available from the corresponding authors upon reasonable request.

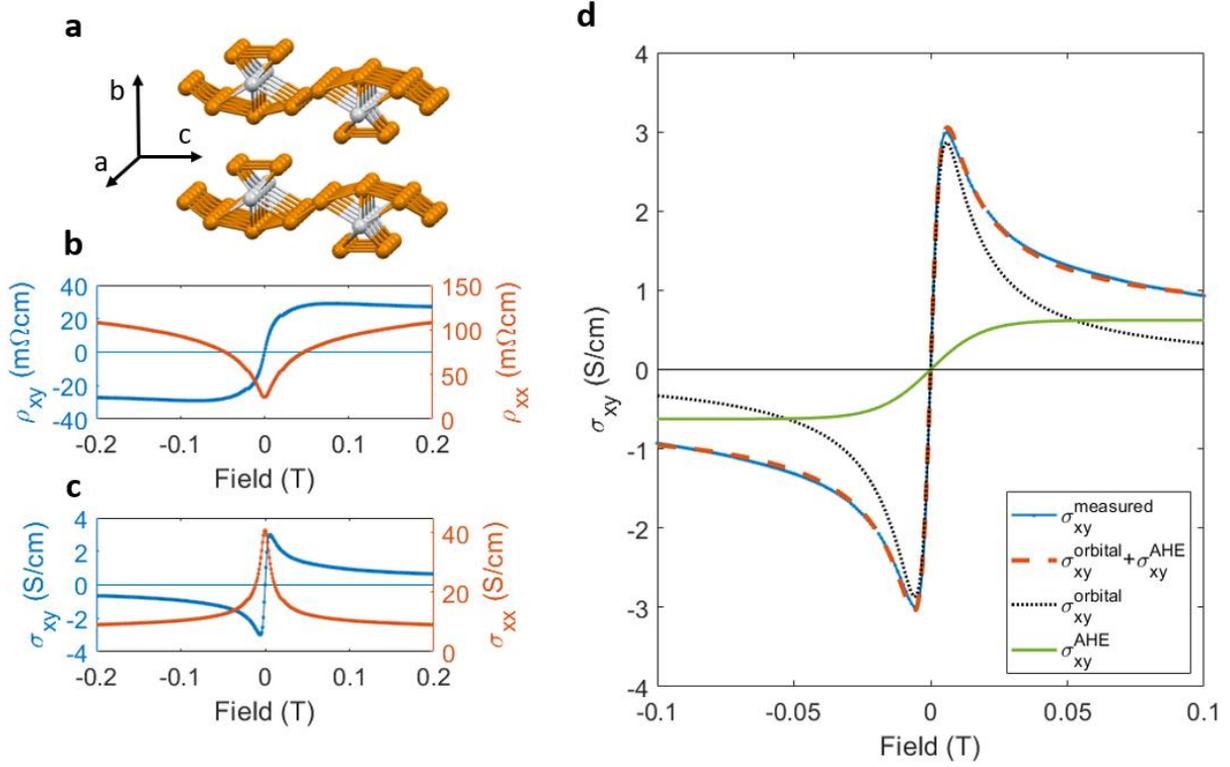

**Figure 1:** (a) Crystal structure of ZrTe$_5$. Chains of tightly bound Zr atoms extend along the *a* lattice direction, forming sheets in the *ac* plane that are van der Waals coupled in the *b* lattice direction. (b) Resistivities $\rho_{xx}$ and $\rho_{xy}$ measured in a bulk crystal at 2K. (c) Conductivities $\sigma_{xx}$ and $\sigma_{xy}$ calculated from $\rho_{xx}$ and $\rho_{xy}$. See Methods for details of the inversion of the resistivity tensor. (d) The Hall conductivity (solid blue) is decomposed into an orbital term $\sigma_{xy}^{orbital} = pq\mu^2/(1+\mu^2 B^2)$ (dotted black) and an anomalous term fitted with a smooth activation function, $\sigma_{xy}^{AHE} = \sigma_0^{AHE}\tanh(B/B_0)$ (solid green). The total fit $\sigma_{xy}^{orbital} + \sigma_{xy}^{AHE}$ is shown in dashed orange.

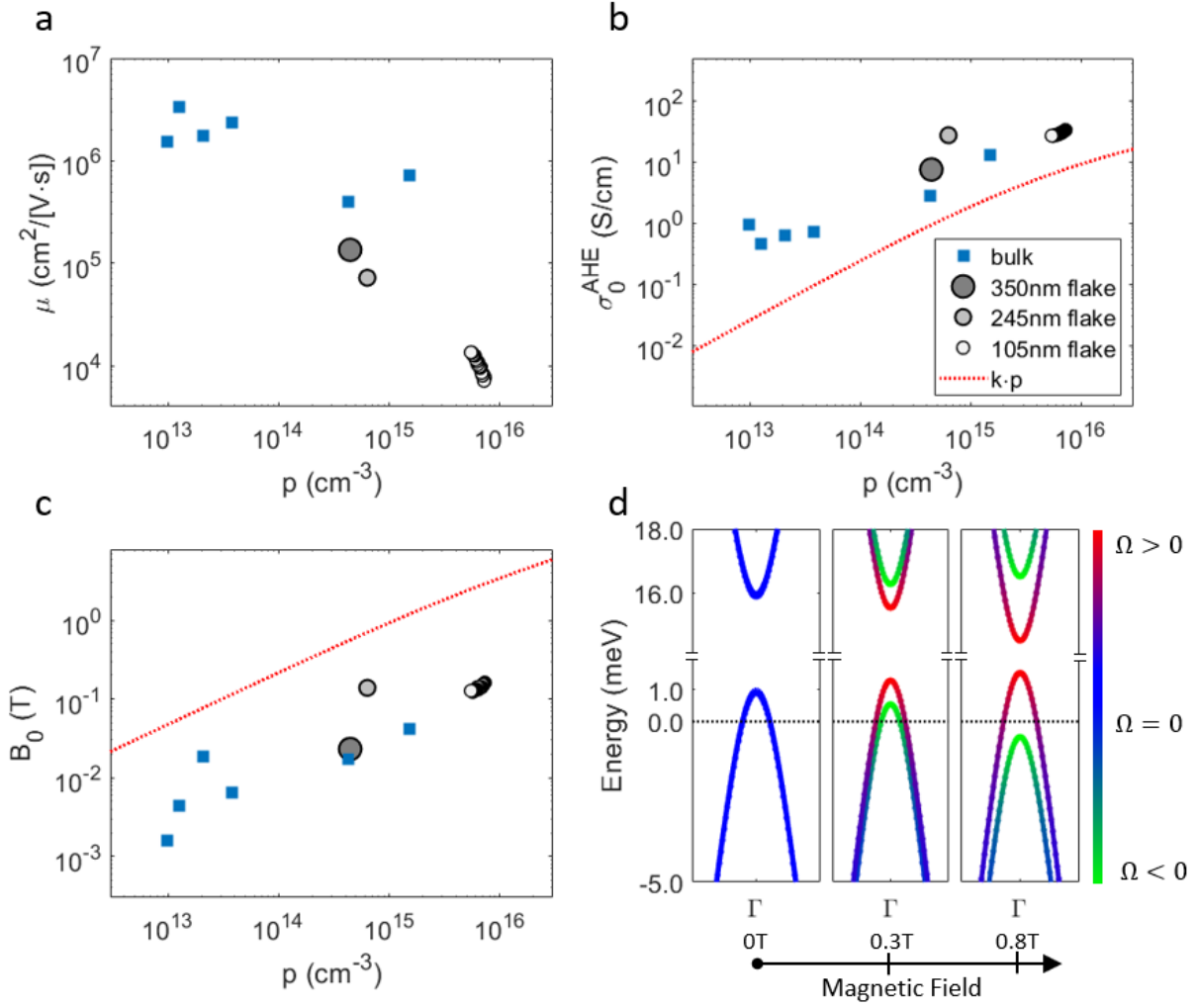

**Figure 2:** (a-c) Mobility (a), saturated anomalous Hall conductivity (b), and saturation field (c) plotted as a function of carrier density. The carrier density is extracted from the fitting procedure of $\sigma_{xy}$ shown in Fig. 1. Blue squares indicate bulk samples. Black circles indicate exfoliated flakes with thicknesses 350 nm, 245 nm, and 105 nm. The 105 nm thick flake was additionally doped with a silicon back gate. The red dotted line in (b) corresponds to the magnitude of $\sigma_0^{AHE}$ for a fully spin polarized band calculated using the $\mathbf{k} \cdot \mathbf{p}$ model. The red line in (c) corresponds the calculated magnetic field strength required to reach full spin polarization. (d) Band structure and Berry curvature of ZrTe$_5$ calculated using the $\mathbf{k} \cdot \mathbf{p}$ model. The Fermi energy corresponds to a carrier density of $5 \times 10^{14}\ cm^{-3}$. The green (red) color denotes the Berry curvature from spin up (down) electrons. The magnetic field is 0, 0.3 and 0.8 T from left to right. At 0.8 T the Zeeman splitting is large enough to fully spin polarize the carriers for the indicated carrier density.

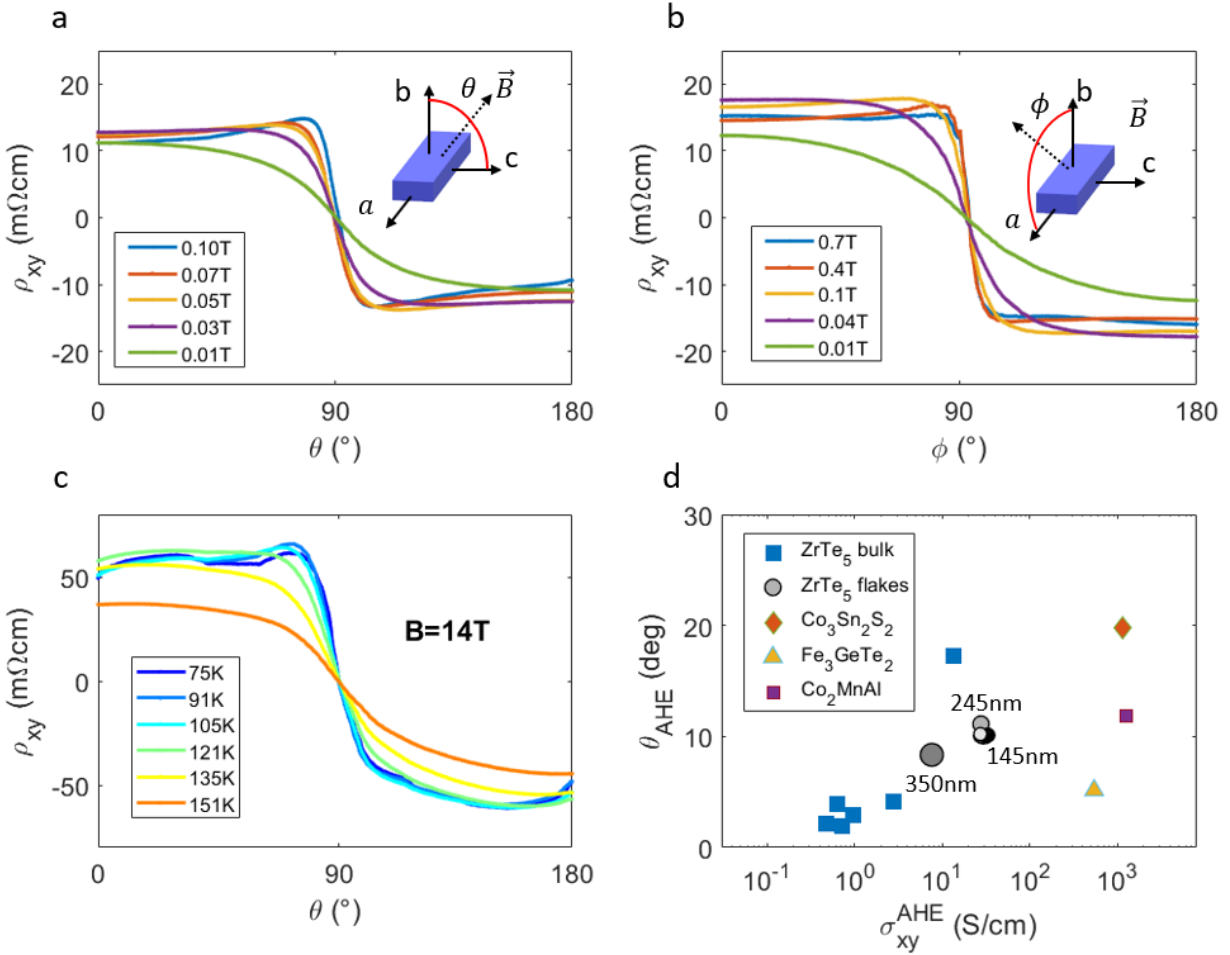

**Figure 3:** (a-b) $\rho_{xy}$ measured at 2K at fixed field, measured as a function of magnetic field angle for a typical low-carrier density bulk crystal. The field is rotated in the *bc* plane in (a), and the *ac* plane in (b), as indicated schematically in the insets. For both orientations of the field, there is a sudden switching of $\rho_{xy}$ as the field crosses the plane. (c) Temperature dependence of $\rho_{xy}$ as the field is rotated across the *bc* plane. The field is fixed at 14 T. A strong deviation from cosine-like behavior is seen up to ~135 K. (d) The anomalous hall angle ($\theta_H^{AHE} = \tan^{-1}(\sigma_0^{AHE}/\sigma_{xx})$) for bulk ZrTe$_5$ (blue squares) and exfoliated flakes of ZrTe$_5$ (black circles). We find $\theta_H^{AHE}$ reaches up to 17.2° in the highest carrier density bulk crystal we measure. For comparison we also show $\theta_H^{AHE}$ for other magnetic topological semimetals.

# Extended Data Figures

# Abrupt switching of the anomalous Hall effect by field-rotation in nonmagnetic ZrTe$_5$

Joshua Mutch[1], Xuetao Ma[2], Chong Wang[3], Paul Malinowski[1], Joss Ayres-Sims[1], Qianni Jiang[1], Zhoayu Liu[1], Di Xiao[3], Matthew Yankowitz[1,2#], Jiun-Haw Chu[1#]

[1]Department of Physics, University of Washington, Seattle, WA 98105, USA
[2]Department of Material Science and Engineering, University of Washington, Seattle, WA 98105, USA
[3]Department of Physics, Carnegie Mellon University, Pittsburg, PA 15213, USA

# correspondence to jhchu@uw.edu; myank@uw.edu

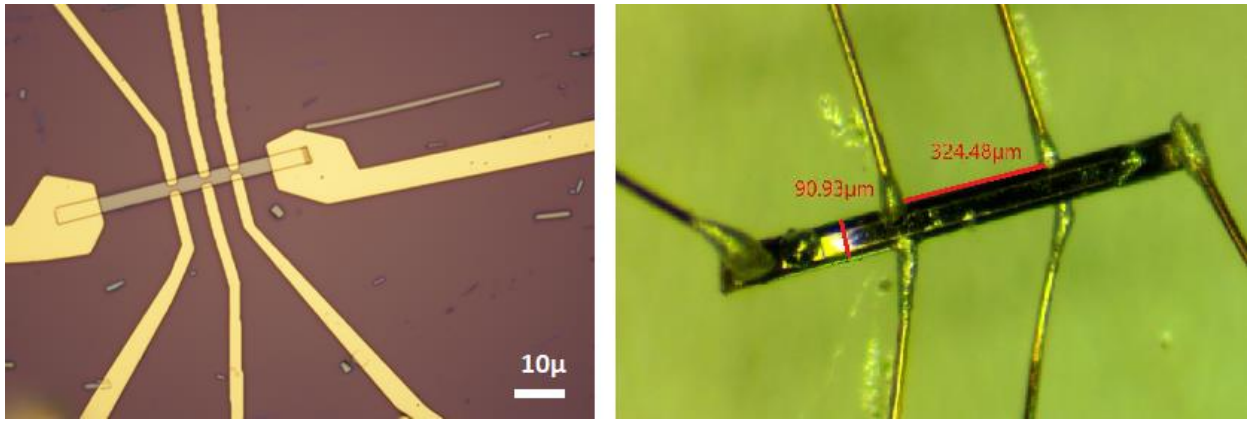

**Extended Data Figure 1:** Images of exfoliated devices and bulk crystals: Left: Exfoliated 245 $nm$ thick flake with gold contacts. Right: Bulk 40 $\mu m$ thick single crystal with 25 $\mu m$ diameter gold wires.

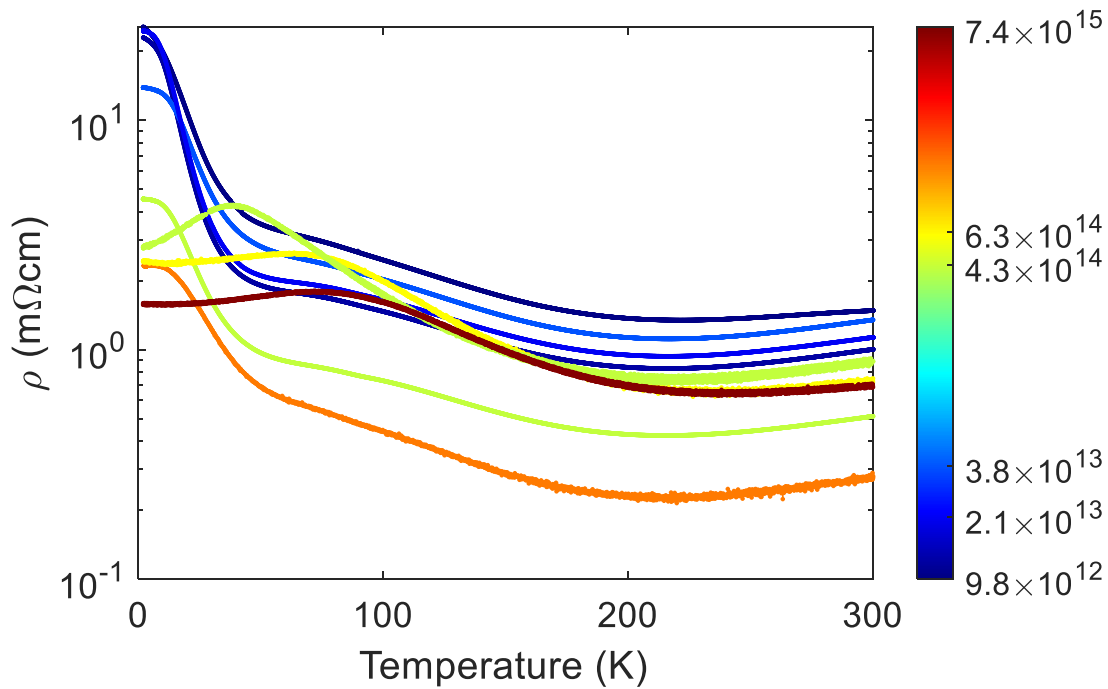

**Extended Data Figure 2**: Resistivity-temperature data for all crystal growth. The color is scaled with the 2K carrier density as derived from $\sigma_{xy}$ fitting described in the main text. The color bar on the right reports the carrier density in units of $cm^{-3}$. Lower carrier density samples measured a higher 2K resistivity.

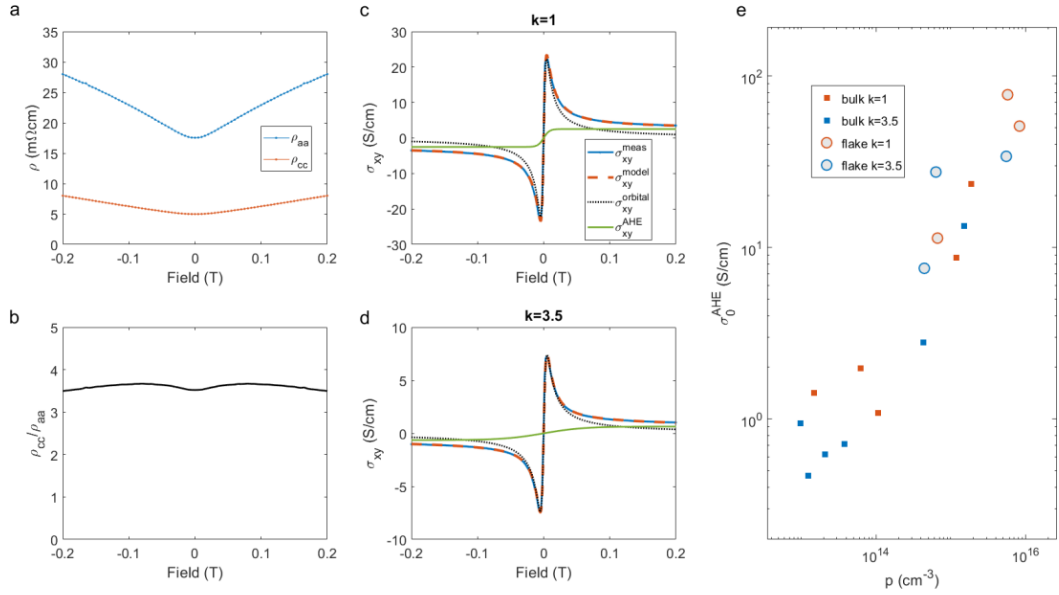

**Extended Data Figure 3:** (a) $\rho_{aa}$ ($\rho_{xx}$) and $\rho_{cc}$ ($\rho_{yy}$) as a function of field at 2K. (b) The resistivity anisotropy ratio $k = \rho_{cc}/\rho_{aa}$ was found to be nearly constant, varying by less than 5% in the relevant field range used in our fitting. (c-d) Hall conductivity and its fitting based on the assumption of $k=1$ (c) and $k=3.5$ (d) when using the formula $\sigma_{xy} = \rho_{xy}/(\rho_{xy}^2 + \rho_{xx}\rho_{yy})$. The magnitude of the total and fitted anomalous Hall conductivities is reduced when $k$ increases. (e) The relationship between $\sigma_0^{AHE}$ and the carrier density $p$ is relatively unchanged when using different values of $k$. Blue data indicates fitting results for $k=1$, and red data for $k=3.5$.

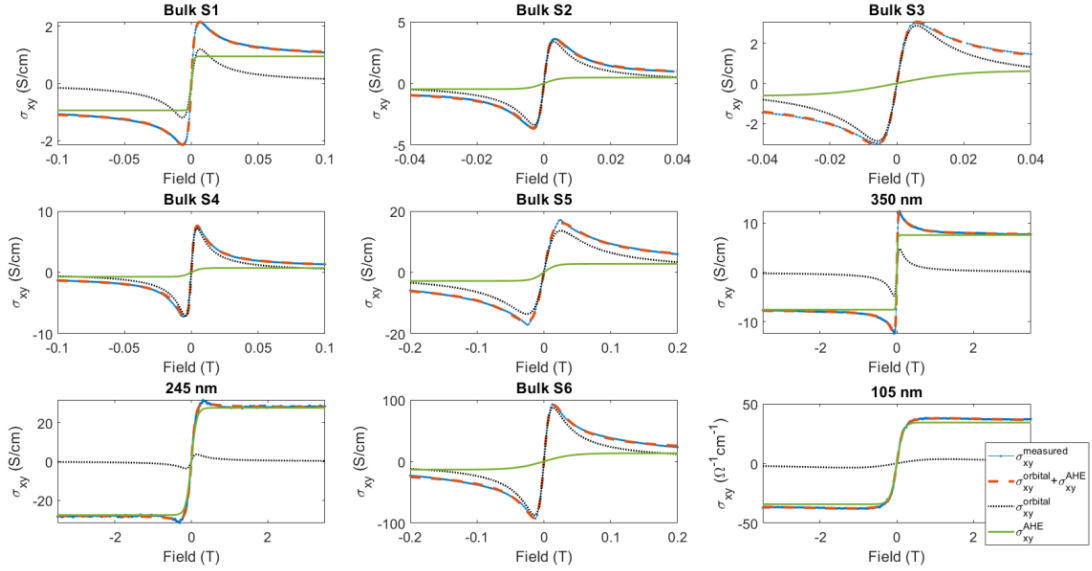

**Extended Data Figure 4:** $\sigma_{xy}$ versus field for all samples measured in this work. The raw data is shown in solid blue, the fitting term $\sigma_{xy}^{orbital} + \sigma_{xy}^{AHE}$ is shown in dashed orange. The individual components $\sigma_{xy}^{orbital}$ and $\sigma_{xy}^{AHE}$ are shown in dotted black and solid green, respectively. Data is presented in order of increasing carrier density from left to right, bottom to top.

| Sample ID | $p\ (cm^{-3})$ | $\mu\ \left(\dfrac{cm^2}{V\cdot s}\right)$ | $\sigma_0^{AHE}\ \left(\dfrac{S}{cm}\right)$ | $B_0\ (T)$ |
|---|---|---|---|---|
| Bulk S1† | $9.8 \times 10^{12}$ | $1.5 \times 10^6$ | 0.95 | 0.01 |
| Bulk S2† | $1.3 \times 10^{13}$ | $3.4 \times 10^6$ | 0.47 | 0.005 |
| Bulk S3† | $2.1 \times 10^{13}$ | $1.7 \times 10^6$ | 0.059 | 0.02 |
| Bulk S4 | $3.8 \times 10^{13}$ | $2.4 \times 10^6$ | 0.72 | 0.007 |
| Bulk S5 | $4.3 \times 10^{14}$ | $4.0 \times 10^5$ | 2.8 | 0.02 |
| 350 nm | $4.4 \times 10^{14}$ | $1.4 \times 10^5$ | 7.6 | 0.023 |
| 245 nm | $6.3 \times 10^{14}$ | $7.2 \times 10^4$ | 27.5 | 0.14 |
| Bulk S6 | $1.5 \times 10^{15}$ | $7.3 \times 10^5$ | 13.3 | 0.04 |
| 105 nm | $7.4 \times 10^{15}$ | $7 \times 10^3$ | 34 | 0.16 |

**Extended Data Table 1**: Carrier density ($p$), mobility ($\mu$) and the saturated anomalous Hall conductivity ($\sigma_0^{AHE}$) extracted from the fitting of $\sigma_{xy}$ for all samples measured in this work. †Grown with arc-melted zirconium.

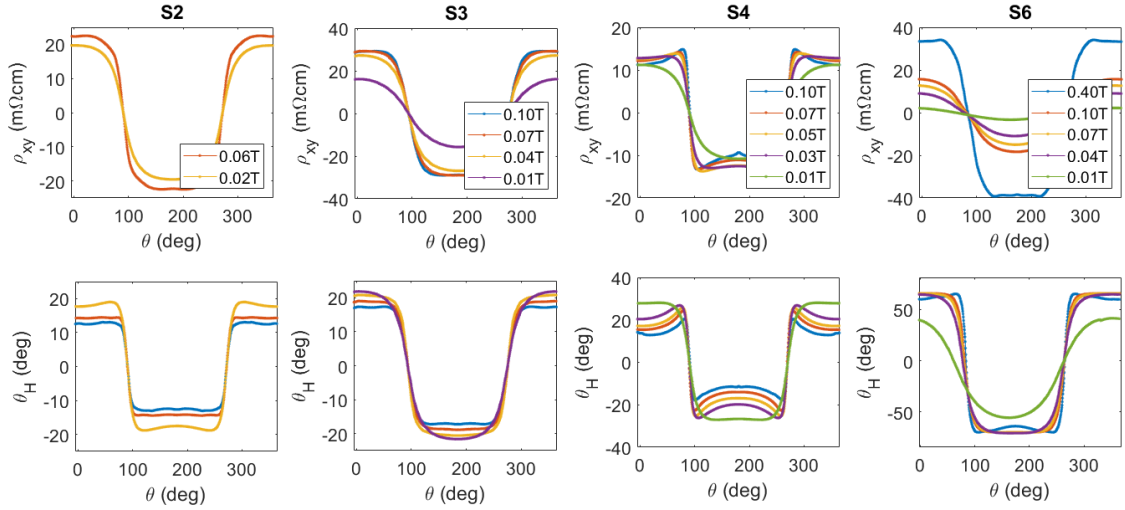

**Extended Data Figure 5**: $\rho_{xy}$ (upper row) and Hall angle $\theta_H$ (bottom row) as a function of magnetic field angle $\theta$ measured from the interlayer direction. The field is rotated in the *bc* plane. Each column represents a different bulk sample.

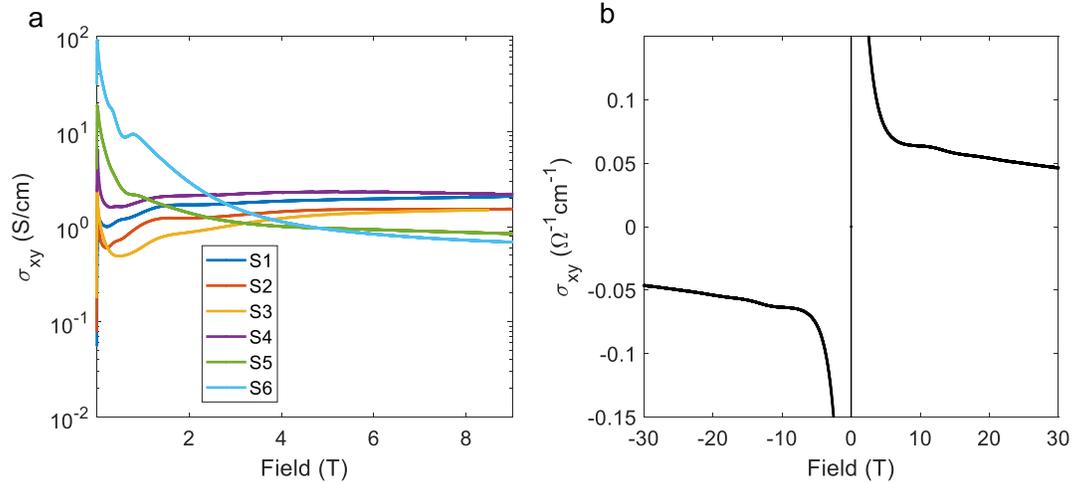

**Extended Data Figure 6:** Saturation of $\sigma_{xy}$ at high fields. (a) Field dependence of $\sigma_{xy}$ up to 9 T for all bulk samples measured in this work. (b) $\sigma_{xy}$ of a representative sample measured up to 30 T.

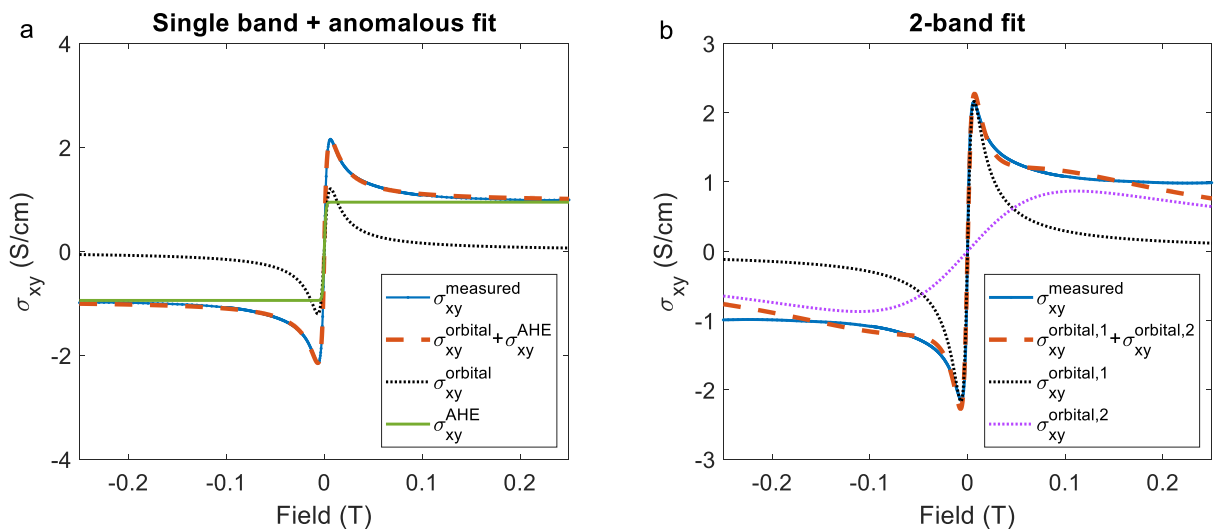

**Extended Data Figure 7:** Comparison of two fitting schemes. (Left) $\sigma_{xy}$ fit with a single band orbital Hall effect and an anomalous Hall term. (Right) Fitting with a 2-band model without an anomalous Hall term.

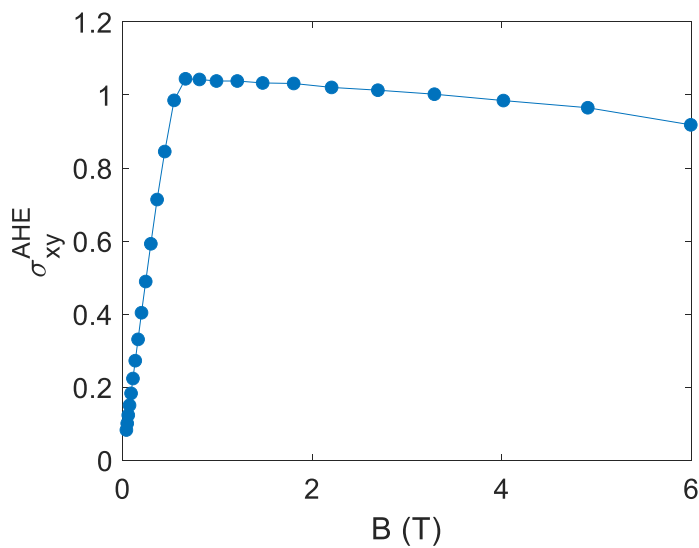

**Extended Data Figure 8:** $\sigma_{xy}$ as a function of magnetic field calculated from the $k \cdot p$ model and assuming a g factor of 21.3 for a carrier density of $5 \times 10^{14} cm^{-3}$.

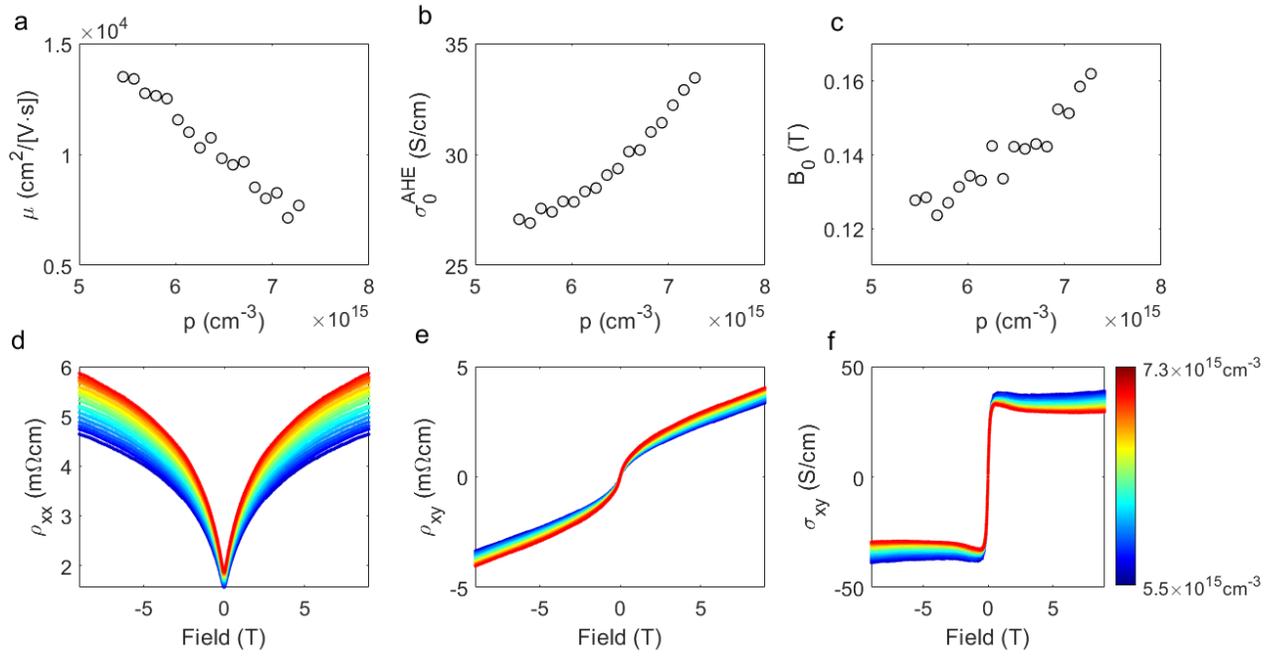

**Extended Data Figure 9:** (a-c) Fitting parameters from the $\sigma_{xy} = \sigma_{xy}^{orbital} + \sigma_{xy}^{AHE}$ model for the gated 105nm flake. (d-f) $\rho_{xx}$, $\rho_{xy}$, and $\sigma_{xy}$ as a function of field for gating voltage setpoints. The color bar denotes the carrier density inferred from the $\sigma_{xy}^{orbital}$ fit.

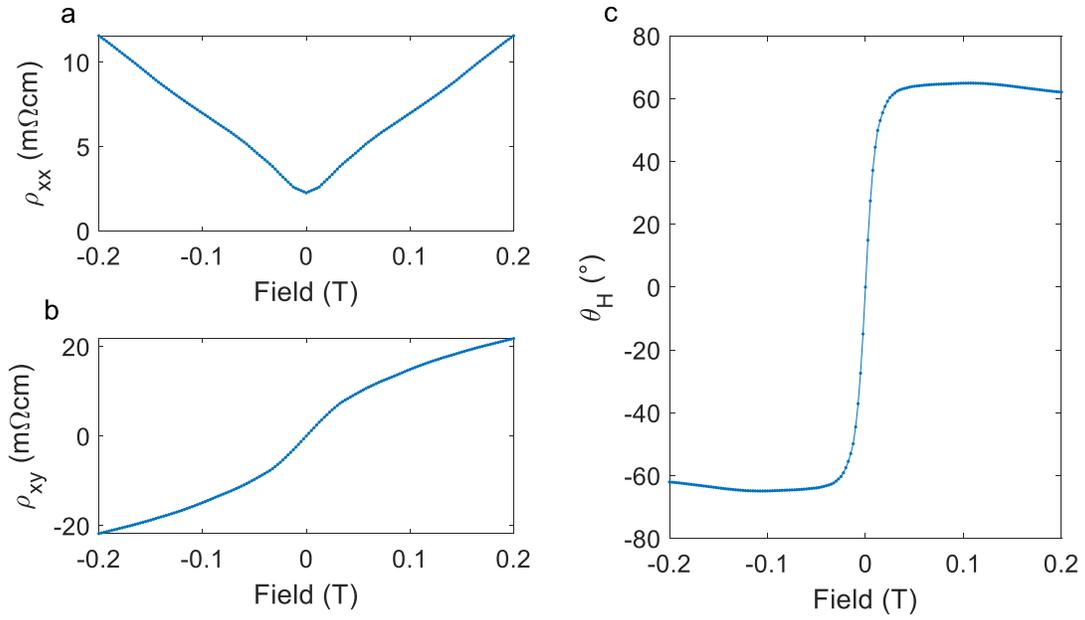

**Extended Data Figure 10:** (a-b) $\rho_{xx}$ and $\rho_{xy}$ at 2K for a bulk crystal with the highest hall angle measured in this work. (c) The hall angle defined as $\theta_H = \tan^{-1}\left(\frac{\rho_{xy}}{\rho_{xx}}\right)$.